\documentclass[runningheads,journal,9pt,lneq]{IEEEtran}
\usepackage{cite}
\ifCLASSINFOpdf
   \usepackage[pdftex]{graphicx}
  \graphicspath{{../pdf/}{../jpeg/}}
  \DeclareGraphicsExtensions{.pdf,.jpeg,.png}
\else
  
   \usepackage[dvips]{graphicx}
   \graphicspath{{../eps/}}
  \DeclareGraphicsExtensions{.eps}
\fi
\usepackage[cmex10]{amsmath}
\usepackage{amsfonts}
\usepackage{amssymb}
\usepackage{mathrsfs}
\usepackage{array}
\usepackage{mdwmath}
\usepackage{mdwtab}
\usepackage{eqparbox}
\usepackage[tight,footnotesize]{subfigure}
\usepackage{fixltx2e}
\usepackage{stfloats}
\usepackage{url}
\newtheorem{theorem}{Theorem}[section]
\newtheorem{definition}{Definition}[section]
\newtheorem{lemma}[theorem]{Lemma}

\newenvironment{proof}[1][Proof]{\begin{trivlist}
\item[\hskip \labelsep {\bfseries #1}]}{\end{trivlist}}

\newcommand{\qed}{\nobreak \ifvmode \relax \else
      \ifdim\lastskip<1.5em \hskip-\lastskip
      \hskip1.5em plus0em minus0.5em \fi \nobreak
      \vrule height0.75em width0.5em depth0.25em\fi}

\newcommand{\AT}{\mathbf{A}_{\xi}}
\newcommand{\Ab}{\mathbf{A}}
\newcommand{\ATa}{\mathbf{A}_{\tau}}
\makeatletter

\newcommand{\Rmnum}[1]{\expandafter\@slowromancap\romannumeral #1@}
\makeatother

\begin{document}
\title{On the Achievability of Cram\'{e}r-Rao Bound\\In Noisy Compressed Sensing}
\author{Rad~Niazadeh,~\IEEEmembership{Member,~IEEE}~Massoud~Babaie-Zadeh,~\IEEEmembership{Senior Member,~IEEE,} and~\\Christian~Jutten,~\IEEEmembership{Fellow,~IEEE}
\thanks{ 
Copyright~\copyright~2011 IEEE. Personal use of this material is permitted. However, permission to use this material for any other purposes must be obtained from the IEEE by sending a request to pubs-permissions@ieee.org. This work has been partially funded by Iran Telecom Research Center (ITRC) and Iran National Science Foundation (INSF). Part of this work was done when the second author was in sabbatical at the University of Minnesota.

R.~Niazadeh is with the Electrical and Computer Engineering Department, Cornell University, Ithaca, NY 14850 (email: rn274@cornell.edu).

 M.~Babaie-Zadeh is with the Electrical Engineering Department, Sharif University of Technology, Tehran 14588-89694, Iran (e-mail:
mbzadeh@yahoo.com).

 C. Jutten is with the GIPSA-Lab, Department of Images and Signals, UMR CNRS 5216, University of Grenoble, Grenoble, France (e-mail: Christian.Jutten@gipsa-lab.grenoble-inp.fr), and with Institut Universitaire de France.}}
\maketitle
\begin{abstract}
%Recently, it has been proved that a joint typical estimator can be used in noisy compressed sensing problem to achieve the Cram\'{e}r-Rao lower bound on %the mean square error. The previous paper around this subject provided a lemma that comprise the main building block of this proof. This mentioned lemma %contains a mathematical mistake in its statement and proof which should be corrected. In this paper, we will explain this mistake and will then %state a new correct form of the lemma. After proving the new version, we will finally restudy the results obtained from this lemma. 
Recently, it has been proved in~\cite{babadi2009asymptotic} that in noisy compressed sensing, a joint typical estimator can asymptotically achieve the Cram\'{e}r-Rao lower bound of the problem. To prove this result, ~\cite{babadi2009asymptotic} used a lemma, which is provided in~\cite{akcakaya711shannon}, that comprises the main building block of the proof. This lemma is based on the assumption of Gaussianity of the measurement matrix and its randomness in the domain of noise. In this correspondence, we generalize the results obtained in~\cite{babadi2009asymptotic} by dropping the Gaussianity assumption on the measurement matrix. In fact, by considering the measurement matrix as a deterministic matrix in our analysis, we find a theorem similar to the main theorem of~\cite{babadi2009asymptotic} for a family of randomly generated (but deterministic in the noise domain) measurement matrices that satisfy a generalized condition known as The Concentration of Measures Inequality. By this, we finally show that under our generalized assumptions, the Cram\'{e}r-Rao bound of the estimation is achievable by using the typical estimator introduced in~\cite{babadi2009asymptotic}.

\end{abstract}
\section{Introduction}
Compressed Sensing (CS) which is also known as Compressive Sampling \cite{donoho2006compressed,candes2006compressive,tsaig2006extensions} is a well known
method for taking linear measurements from a sparse vector. Compressed sensing proposes that one can recover a sparse signal from a few number of measurements, and so it can override the usual sampling method based on Nyquist criteria\cite{donoho2006compressed}. In this correspondence, we revisit the problem of signal recovery in noisy compressed sensing, in which the above mentioned measurements are blended with noise. Indeed, suppose that noisy measurements of the sparse signal are taken by a random measurement matrix in the following form:
\begin{equation}
\label{eq:ncs}
\mathbf{y}={\Ab}\mathbf{s}+\mathbf{n}~,
\end{equation} 
in which, $\mathbf{s}$ is the original ${M\times 1}$ sparse signal, $\mathbf{y}$ is the ${N\times 1}$ vector of measurements, $\mathbf{n}\sim N(0,\sigma_{n}^2I_{N\times N})$ is an ${N\times 1}$ Gaussian noise vector and $\mathbf{A}=[\mathbf{a}_1~~\mathbf{a}_2\dots\mathbf{a}_M]$ is an ${N\times M}$ measurement matrix whose elements are usually generated at random. More precisely, these elements are independent and identically distributed random variables drawn from some specific distributions (such as Gaussian, Bernoulli, etc.), so that the overall measurement  matrix will be appropriate in the framework of recovery in compressive sampling~\cite{donoho2006compressed,candes2006robust,candes2006compressive,baraniuk2008compressive}.
% More precisely, these elements are independent and identically distributed generated random variables drawn from $N(0,1)$.
Suppose that $\mathbf{s}$ is sparse, i.e. $\lVert\mathbf{s}\rVert_{0}=K\ll M$ where $\lVert.\rVert_0$
denotes the $l_0$-norm, i.e. the number of non-zero components of $\mathbf{s}$. Moreover, define $\tau\triangleq \textrm{supp}(\mathbf{s})$ as a subset of $\{1,2\dots M\}$ that contains the indices of non-zero elements of $\mathbf{s}$, i.e. $\tau=\big\{i \in \{1,2,\dots M\}:s_i\neq 0\big\}$ in which $s_i$ stands for the $i$th element of $\mathbf{s}$. For this model, one can also define the size parameters~\footnote{In the context of compressive sampling, the linear system in (\ref{eq:ncs}) is under-determined, i.e. $M>N$. However, this assumption is not required in any of our presented analyses. Hence, our provided lemmas and theorems in this correspondence could be applied to the case of overdetermined noisy sparse recovery, which appears in many applications in communication theory, for example in sparse channel estimation~\cite{niazadeh2010alternating}.} as in \cite{babadi2009asymptotic}:
\begin{equation}
\label{eq:sp1}
\alpha\triangleq\frac{K}{N} ~~~\beta\triangleq\frac{M}{K}~.
\end{equation}

The main problem of compressive sampling is to estimate the unknown sparse signal from its noisy measurements which are taken as in (\ref{eq:ncs}). Many efforts have been done to find a practical recovery method and some acceptable solutions have been proposed in the literature whose computational cost are tolerable, such as the algorithms that are proposed in~\cite{candes2006stable,tropp2007signal,donoho2006most,chen2001atomic,blumensath2008gradient,gorodnitsky1997sparse,needell2009cosamp,mohimani2009fast}.
On the other hand, there is another related problem which is indeed the framework of our correspondence. In this problem, we are searching for the existence of an \textit{efficient} estimator, an estimator that can achieve the Cram\'{e}r-Rao lower bound~\cite{kay1993fundamentals} for the Mean Square Error~(MSE) of the estimation. It is important to note that in this problem, computational complexity of the proposed estimator has no importance (or much less importance when comparing to practical methods), while the achievablity of Cram\'{e}r-Rao bound and the \textit{existence} of such an estimator is in the point of interest.

For our problem, two different Cram\'{e}r-Rao lower bounds for MSE have been studied in~\cite{babadi2009asymptotic,carbonelli2007sparse} depending on the amount of knowledge of the estimators about the sparsity structure of the original vector.
The first bound, which is known as CRB-S~\cite{carbonelli2007sparse}, is the Cram\'{e}r-Rao lower bound of a Genie Aided Estimation (GAE) problem in which the estimators know the location of the non-zero taps i.e. $\tau$, as if a
Genie has aided them with the location of the taps~\cite{candes2007dantzig,carbonelli2007sparse}. This bound can be described in closed form as~\cite{carbonelli2007sparse}:
\begin{equation}
\label{eq:CRBcompare}
\textrm{CRB-S}=\sigma_n^2 \textrm{Trace}\{(\mathbf{A}_\tau^T\mathbf{A}_\tau)^{-1}\}~,
\end{equation}  
in which $\mathbf{A}_\tau$ is a sub-matrix of $\mathbf{A}$ that contains the columns corresponding to the indices in $\tau$. Among all of the estimators that know the location of the taps, it can be shown that (as we will also show later in this correspondence) the efficient estimator will be the Structural Least Square Estimator (SLSE) which finds the solution of the following problem~\cite{carbonelli2007sparse}:
\begin{equation}
\label{eq:snos}
\mathbf{\hat{s}_\tau}=\underset{\mathbf{s_\tau}}{\textrm{argmin}}\lVert\mathbf{y}-\mathbf{\mathbf{A}_\tau}\mathbf{s_\tau}\rVert_2^2\enspace ~,
\end{equation}
in which $\mathbf{s}_\tau$ is the $K\times 1$ vector of non-zero taps. The second bound, which is known as CRB-US~\cite{babadi2009asymptotic}, is the Cram\'{e}r-Rao lower bound for the estimation problem in which the estimators have only prior knowledge about the cardinality of $\tau$ i.e. $K$, which indicates the degree of sparsity. 
It is obvious that the Cram\'{e}r-Rao bound for this kind of estimation is not less than that of GAE, i.e.:
\begin{equation}
\label{eq:neq}
\textrm{CRB-US}\geq\textrm{CRB-S}~.
\end{equation}
Furthermore, in a recent work by Ben-Haim et al.~\cite{ben2010cramér} an expression for CRB-US has been stated. In fact, they have shown that the behaviour of the CRB differs depending on whether or not the unknown sparse vector has maximal support (i.e., $\lVert \mathbf{s}\rVert_0=K$ or $\lVert \mathbf{s}\rVert_0<K$). More accurately, they have shown that if the measurement matrix satisfies the uniqueness theorem provided by Donoho et al.\cite{donoho2006compressed} and Cand\'es et al.\cite{candes2006compressive}, and if we consider the case of maximal support, i.e. when $\lVert \mathbf{s}\rVert_0=K$ which is indeed our case in this correspondence, and if we consider the case of finite size sparse recovery, i.e. when $M$, $N$, and $K$ are fixed and limited, the Cram\'{e}r-Rao bound equals to that of  GAE (when the sparsity pattern is known by the estimator), i.e. CRB-US equals CRB-S. However, according to our best knowledge, no evidence of exact achievability of CRB-US by the means of any practical estimator or non-practical estimators has been presented in the literature for the case of fixed and limited $M$, $N$, and $K$. So, if someone proposes an estimator that can achieve CRB-S instead of CRB-US while it has only prior knowledge about the sparsity degree, then it will be proven that CRB-S and CRB-US are equal to each other (as stated in~\cite{ben2010cramér}) and both of them are achievable by this proposed estimator.

Many efforts have been done to design an estimator with just the knowledge about the cardinality of $\tau$ that can achieve MSE as close as possible to the GAE Cram\'{e}r-Rao lower bound (CRB-S). Cand\'es et al.\cite{candes2007dantzig} and Haupt et al.\cite{haupt2006signal} proposed estimators that can achieve CRB-S up to a factor of $\log M$ which is far from CRB-S. Interestingly, recent works done by Babadi et al.\cite{babadi2009asymptotic} and Ak\c cakaya et al.\cite{akcakaya711shannon} have shown that by using an impractical estimator known as ``Typical Estimator'', under certain constraints on $\mathbf{s}$ and $\mathbf{A}$, one can asymptotically achieve the Cram\'{e}r-Rao bound of the GAE problem, i.e. CRB-S, without a priori knowing $\tau$. By asymptotic, we mean where $N$, $M$ and $K$ tend to infinity while the size parameters in~(\ref{eq:sp1}) remain constant. In other words, since the proposed typical estimator asymptotically achieves CRB-S, one can conclude that \textbf{a)} CRB-S and CRB-US are asymptotically equal, and \textbf{b)} this Cram\'{e}r-Rao bound is achievable (note that in general, the  Cram\'{e}r-Rao bound of an estimation problem is not achievable, i.e. it is not generally a tight bound for MSE).

  The typical estimation in~\cite{akcakaya711shannon,babadi2009asymptotic} is based on checking the \textit{Joint Typicality} of the noisy observations vector with all possible choices of $\tau$, and then decoding the one which is jointly typical with the observed $\mathbf{y}$. Definition of joint typicality is introduced in \cite{akcakaya711shannon,babadi2009asymptotic} and we will review it later in this correspondence.\footnote{It is worth noting that the concepts of typicality and typical estimation have been first introduced in the literature of Shannon's work on information theory\cite{cover2006elements,shannon2001mathematical}. With some changes, this concept is adapted to the field of compressive sampling in\cite{akcakaya711shannon,babadi2009asymptotic}.} After detecting the support of $\mathbf{s}$, typical estimator estimates the unknown vector $\mathbf{s}$ by using a structural least square estimation method, i.e. it finds the solution of (\ref{eq:snos}). In \cite{babadi2009asymptotic}, the proof of the achievability of the Cram\'{e}r-Rao bound by using the typical estimator is based on a lemma (Lemma 3.3 of \cite{akcakaya711shannon}), which bounds the probability of two error events in the mentioned estimation process. The first of these probabilities is the probability of the event that the support of $\mathbf{s}$ is not jointly typical with $\mathbf{y}$ which we denote\footnote{We will use the notations ``$\sim$'' and ``$\nsim$'' for indicating a jointly typical or a non jointly typical pair in the rest of this correspondence.} by $(\tau \nsim \mathbf{y})$ and the second one is the probability of the event that a subset $J\subset\{1,2,\dots M\}\neq\tau$ with cardinality $K$ is jointly typical with $\mathbf{y}$ which we denote by $(J \sim \mathbf{y})$. Using this lemma, \cite{babadi2009asymptotic} shows that if the average power of $\mathbf{s}$ is limited and if
\begin{equation} 
\label{babadi_region}
\alpha<\frac{1}{9+4\log(\beta-1)}~,
\end{equation}
then the joint typical estimator achieves the Cram\'{e}r-Rao bound as $N\rightarrow\infty$. 

It is important to mention that the proof of the above mentioned statement in~\cite{babadi2009asymptotic} depends on the assumption that the elements of the measurement matrix are drawn randomly from a Gaussian distribution, in addition to the assumption that this matrix is stochastic in the noise domain. By this, we mean that this assumption will impose the consideration of the elements of measurement matrix as random variables in our analysis, just like the elements of noise vector. On the contrary, these assumptions are unnecessary in the ordinary framework of compressed sensing, while we are looking to find a stable recovery method. In fact, it is common to use \textit{non-stochastic} but \textit{randomly generated} measurement matrices in this context, while assuming that the noise vector is stochastic (because the estimator knows the exact measurement matrix, but it is not aware of the noise vector). Additionally, among all randomly generated matrices, appropriate measurement matrices are those that satisfy a constraint called \textit{ The Concentration of Measures Inequality}\footnote{This condition is a preliminary condition for Restricted Isometry Property (RIP) which is a well-known sufficient condition in the area of compressed sensing for robust and stable recovery of the original sparse vector via $l_1$-minimization~\cite{candes2006robust,candes2006stable}.}, i.e. the following condition~\cite{baraniuk2008simple}:
\begin{equation}
\label{con:eq}
\mathbb{P}\{\lvert~\lVert\mathbf{A}\mathbf{x}\rVert^2-\lVert \mathbf{x}\rVert^2\rvert\geq \epsilon\lVert \mathbf{x}\rVert^2\rvert \}\leq 2e^{-N c_0(\epsilon)},~\epsilon \in (0,1)~~~~~~~~,
\end{equation}
where the probability is taken over random space for $N \times M$ random-generated matrix $\mathbf{A}$, $\epsilon\in(0,1)$ is arbitrary, $c_0(\epsilon)$ is a constant depending only on $\epsilon$ and such that for all $\epsilon\in(0,1)$, $c_0(\epsilon)>0$ and $\mathbf{x}$ is an arbitrary fixed vector in $\mathbb{R}^M$. 
  Because of this mentioned difference in the assumptions made in~\cite{babadi2009asymptotic} and ordinary assumptions made in the framework of compressed sensing, one may wonder that the results obtained in~\cite{babadi2009asymptotic} may be also valid in the case of a larger family of measurement matrices than just the Gaussian matrices. Indeed, we will introduce a family of random-generated matrices which satisfies a modified version of concentration of measures inequality, i.e. the following condition: 
\begin{equation}
\label{con2:eq}
\mathbb{P}\{\lvert~\lVert\mathbf{A}\mathbf{x}\rVert^2-N\lVert \mathbf{x}\rVert^2\rvert\geq \epsilon N\lVert \mathbf{x}\rVert^2\rvert \}\leq 2e^{-N c_0(\epsilon)},~\epsilon \in (0,1)~~~~~~~~,
\end{equation}
in which all the variables are the same as those in~(\ref{con:eq}). Perhaps, the most prominent example of matrices that satisfy (\ref{con2:eq}), are those with elements drawn independently and identically distributed according to $N(0,1)$~\cite{baraniuk2008simple}; but, there is no force on having Gaussian entries in the measurement matrix. More precisely, one can also use matrices whose entries are independent realizations of $\pm 1$ Bernoulli random variables 
\begin{equation}
\label{dis1}
A_{i,j} = \left\{
\begin{array}{rl}
+1 & \text{with probability } 1/2,\\
-1 & \text{with probability } 1/2.
\end{array} \right.
\end{equation}
or related distributions such as
\begin{equation}
\label{dis2}
A_{i,j} = \left\{
\begin{array}{rl}
+\sqrt{3} & \text{with probability } \frac{1}{6},\\
0 & \text{with probability } \frac{2}{3},\\
-\sqrt{3} & \text{with probability } \frac{1}{6}.
\end{array} \right.
\end{equation}
and yet these matrices satisfy (\ref{con2:eq}). In addition to example random matrices described in (\ref{dis1}) and (\ref{dis2}), there are many other examples of random matrices that satisfy the condition in (\ref{con2:eq}) and have an important role in statistical signal processing, communications\footnote{Many applications of using such non-Gaussian random projection, such as sparse channel estimation\cite{niazadeh2010alternating}, have been reported in the literature.}, and in particular compressive sampling. In fact, there is a well known class of linear projections, mostly known as \textit{data-base friendly} random projections~\cite{achlioptas2003database}, that satisfies the condition in (\ref{con2:eq}), and at the same time can exploit the full allotment of dimensionality of a high-dimensional point set. Random i.i.d Gaussian matrices and those in (\ref{dis1}) and (\ref{dis2}) are considered as examples within this class. Hence, as satisfying (\ref{con2:eq}) is a general property of commonly used random projection in signal processing and compressive sampling, it may be interesting to generalize the results obtained in~\cite{babadi2009asymptotic} for this class of matrices. Then, one can conclude that the Cram\'{e}r-Rao bound of the estimation is also asymptotically achievable by using the typical estimator introduced in~\cite{babadi2009asymptotic} and~\cite{akcakaya711shannon}, while we use non-Gaussian matrices that satisfy the condition depicted in (\ref{con2:eq}), which is a common and general condition for measurement matrices in compressed sensing according to the literature. 
%In this paper, we will first point out that there is a mathematical mistake in Lemma 3.3 of \cite{ak�akaya711shannon} in both its statement and its proof. More precisely, we will show that the probability
% on joint typicality obtained in Lemma 3.3 of \cite{ak�akaya711shannon} are not correct. Since this lemma has been used as the main building block to obtain the results of \cite{babadi2009asymptotic}, one wonders that those results (achievablity of CRB-S and asymptotic equivalence of CRB-S and CRB-US)%may also be incorrect and hence they should be revisited. In this paper, after stating the mathematical problem of Lemma 3.3 in \cite{ak�akaya711shannon}, we state its correct form. We will see that although we prove the new form of the theorem in a completely different way compared to the original one (by using Chernof tail bounds for probability), the final result will have very minor differences from the form stated in ^\cite{ak�akaya711shannon}. Then based on the correct form of this theorem, we re-study the results of \cite{babadi2009asymptotic}. We will finally see that although the main theorem used in \cite{babadi2009asymptotic} has been changed, fortunately, all of the results in \cite{babadi2009asymptotic} remain valid, that is, in noisy compressed sensing, the Cram\'{e}r-Rao bound is asymptomatically achievable by using a typical estimator described in \cite{babadi2009asymptotic}, and the constraint in (\ref{babadi_region}) will also be valid without any changes.     

  In this correspondence, according to the above discussion, we investigate the results obtained in~\cite{babadi2009asymptotic}, and then we generalize the conditions for the problem of asymptotic achievability of Cram\'{e}r-Rao bound in noisy compressed sensing. More accurately, by using an alternative approach to this problem comparing to the one used in~\cite{babadi2009asymptotic} and~\cite{akcakaya711shannon}, i.e. by assuming that the measurement matrix, $\mathbf{A}$, is not stochastic in the noise domain, we will find a lemma similar to Lemma 3.3 in \cite{akcakaya711shannon} and prove it using a different method compared to the original one (by using Chernof tail bounds for probability~\cite{hagerup1990guided}). Since Lemma 3.3 of~\cite{akcakaya711shannon} has been used as the main building block to obtain the results of \cite{babadi2009asymptotic}, one wonders if those results (achievablity of CRB-S and asymptotic equivalence of CRB-S and CRB-US) may be incorrect under our new assumption ($\mathbf{A}$ is just generated at random, but it is deterministic when compared to noise) and hence if they should be revised. In this purpose, we first re-state our proved lemma in the case of randomly generated (but deterministic in noise domain) measurement matrices that satisfy (\ref{con2:eq}). Subsequently, we see that the final obtained form have very minor differences from Lemma 3.3 in~\cite{akcakaya711shannon}, while it is valid under the assumption that $\mathbf{A}$ is a deterministic randomly generated matrix. Finally, we re-study the results of~\cite{babadi2009asymptotic} and see that although the main lemma used in~\cite{babadi2009asymptotic} has been changed in our analysis, fortunately, all of the results in \cite{babadi2009asymptotic} remain valid. In other words, in noisy compressed sensing and under our modified version of concentration of measures inequality condition, the Cram\'{e}r-Rao bound is asymptomatically achievable by using a typical estimator described in \cite{babadi2009asymptotic}, and the constraint in (\ref{babadi_region}) will also be valid without any changes.

   This correspondence paper is organized as follows. In the next section, we will first review the definition of joint typicality and the typical estimator introduced in~\cite{babadi2009asymptotic}. Moreover, the main theorem of~\cite{babadi2009asymptotic} and the Lemma 3.3 in~\cite{akcakaya711shannon} will be re-studied. Indeed, we provide a new form of the mentioned lemma under our new assumptions, in which the measurement  matrix is considered as a randomly generated matrix that satisfies (\ref{con2:eq}), although is deterministic in the noise domain. In Section \Rmnum{3}, the Cram\'{e}r-Rao lower bound on MSE for the compressed sensing problem in a noisy setting will be discussed and we will show that the results obtained in~\cite{babadi2009asymptotic} remain valid under our generalized assumptions. So the  Cram\'{e}r-Rao bound of the GAE problem and that of the problem in which estimators have only prior knowledge about the degree of sparsity are asymptotically equal if the measurement matrix satisfies (\ref{con2:eq}), although it may  not be Gaussian or random in the noise domain. In all of the above discussions, we will use the model described in (\ref{eq:ncs}) and we will assume that the matrix $\mathbf{A}$ is randomly generated, but since it is \textit{known} to the estimator, it should be treated as a deterministic matrix.

\section{Statement and proof of the main theorem}
First, consider the noisy compressed sensing model in (\ref{eq:ncs}). As in~\cite{babadi2009asymptotic}, we use the following definition for joint typicality: 
\begin{definition}[Joint Typicality]
Suppose that $\xi\subset\{1,2,\dots M\}$ and $\lvert \xi\rvert=K$, in which $\lvert \cdot\rvert$ denotes the cardinality of a set. Let $\mathbf{A}_{\xi}$ denote the $N\times K$ sub-matrix of $\mathbf{A}$ including those columns of $\mathbf{A}$ that correspond to the indices in $\xi$. Let also $\Pi_{\mathbf{A}_{\xi}}\triangleq\AT(\AT^T\AT)^{-1}\AT^T$ and $\Pi_{\mathbf{A}_{\xi}}^{\bot}\triangleq \mathbf{I} -\Pi_{\mathbf{A}_{\xi}}$. $\xi$ and $\mathbf{y}$ are said to be jointly typical with order $\epsilon$, denoted by ($y\sim\xi)_{\epsilon}$, if and only if:
\begin{equation}
\lvert \frac{1}{N}\lVert\Pi_{\mathbf{A}_{\xi}}^{\bot}\mathbf{y}\rVert ^2-\frac{N-K}{N}\sigma_{n}^2\rvert<\epsilon~.
\end{equation}
\end{definition} 

In order to generalize the results in~\cite{babadi2009asymptotic}, we neglect the assumption that $\mathbf{A}$ is a Gaussian random matrix in the noise domain. Indeed, we assume that $\mathbf{A}$ is a randomly generated matrix, but is known to the estimator, and hence should be considered as a deterministic matrix. Accordingly, we first introduce the following theorem, which is similar to Lemma 3.3 in~\cite{akcakaya711shannon} and only depends on our new assumption on measurement matrix:
 \begin{theorem}[Bounds on the Probabilities of Typicality] 
\label{maint}
 Assume that in (\ref{eq:ncs}), $\tau=\textrm{supp}(\mathbf{s})$. Additionally, assume that $\xi\subset\{1,2,\dots M\}$ and $\lvert \xi\rvert=K$. Considering an arbitrary small enough $\epsilon>0$, the following expressions hold as $N\to\infty$:

\begin{align}
\label{th1}
&\mathbb{P} \{\lvert \frac{1}{N}\lVert\Pi_{\mathbf{A}_{\tau}}^{\bot}\mathbf{y}\rVert ^2-\frac{N-K}{N}\sigma_{n}^2\rvert>\epsilon  \}\stackrel{\text{exponentially}}{\longrightarrow}  0~~, \\ 
&\mathbb{P}\{\lvert \frac{1}{N}\lVert\Pi_{\mathbf{A}_{\xi}}^{\bot}\mathbf{y}\rVert ^2-\frac{N-K}{N}\sigma_{n}^2\rvert<\epsilon  \}\leq\nonumber \\
\label{th2}
&\exp\big\{-\frac{N-K}{4}\Big[\frac{\frac{1}{N}\sum_{i\in\tau\backslash\xi}{\sum_{j\in\tau\backslash\xi}{s_i s_j \mathbf{a'}_i^T\mathbf{a'}_j}}-\epsilon}{\frac{2}{N}\sum_{i\in\tau\backslash\xi}{\sum_{j\in\tau\backslash\xi}{s_i s_j \mathbf{a'}_i^T\mathbf{a'}_j}}+{\sigma'}_n^{2}} 
\Big]^2       \big\}~~.
\end{align}
in which ${\sigma'}_n^{2}=(1-\alpha)\sigma_n^2$ and $\mathbf{a'}_i=(\mathbf{U}_{\xi}\mathbf{a}_i)_{N-K}$, where $\mathbf{U}_{\xi}$ is a unitary matrix extracted from the eigenvalue decomposition of $\Pi_{\AT}^{\bot}$, i.e. $\Pi_{\AT}^{\bot}=\mathbf{U}_{\xi}\mathbf{D}\mathbf{U}_{\xi}^T$ and $\mathbf{D}$ is a diagonal matrix. The $(.)_{m}$ operator denotes a vector comprising of the first $m$ elements of the operand.
\begin{proof}[\textbf{Proof of (\ref{th1})} ]
The proof of this part is the same as the proof of the first part of Lemma 3.3 in \cite{akcakaya711shannon} with some minor modifications. For the sake of readability, we will go through the steps of this proof. In these steps, we will try to find the PDF (Probability Density Function) of $\lVert\Pi_{\mathbf{A}_{\tau}}^{\bot}\mathbf{y}\rVert ^2$ assuming that $\mathbf{A}$ is \textit{known} and deterministic, while the noise vector is random.

Due to the fact that $\Pi_{\mathbf{A}_{\tau}}$ is the projector transform onto $\cal{S}=\textrm{span}\{\text{columns of}~\ATa\}$ and since $\mbox{supp}(\mathbf{s}) = \tau$, we have:
\begin{equation}
\Pi_{\mathbf{A}_{\tau}}^{\bot}\mathbf{y}=\Pi_{\mathbf{A}_{\tau}}^{\bot}(\ATa\mathbf{s}_{\tau}+\,\mathbf{n})=\mathbf{0}+\Pi_{\mathbf{A}_{\tau}}^{\bot}\mathbf{n}\nonumber~.
\end{equation}
 $\Pi_{\mathbf{A}_{\tau}}^{\bot}$ is a symmetric matrix, therefore we can decompose it as $\mathbf{U}_{\tau}\mathbf{D}\mathbf{U}_{\tau}^T$, in which 
$\mathbf{D}$ is a diagonal matrix and $\mathbf{U}_\tau$ is a unitary matrix ($\mathbf{U}_\tau \mathbf{U}_\tau^T=\mathbf{I}$). $\Pi_{\mathbf{A}_{\tau}}^{\bot}$ is an $N\times N$ matrix (which obviously has $N$ eigenvalues). In addition to that, \cite{babadi2009asymptotic} shows that $\ATa$ is full-rank with probability $1$. This means that $\cal{S}=\textrm{span}\{\text{columns of}~\ATa\}$ is a $K$ dimensional subspace of $\mathbb{R}^N$ as $N\rightarrow\infty$. Moreover, for every $\mathbf{y}\in\cal{S}$, we have $\Pi_{\mathbf{A}_{\tau}}^{\bot}\mathbf{y}=0$ and so, the $K$ basis vectors of $\cal{S}$ are $K$ linearly independent eigenvectors of $\Pi_{\mathbf{A}_{\tau}}^{\bot}$ corresponding to the eigenvalue $0$. Additionally, for every $\mathbf{y}\in\cal{S}'=\{\textrm{orthogonal compliment of $\cal{S}$ in $\mathbb{R}^N$} \}$, we have $\Pi_{\mathbf{A}_{\tau}}^{\bot}\mathbf{y}=\mathbf{y}$. In a similar way, we can show that $\cal{S}'$ is an $N-K$ dimensional subspace of $\mathbb{R}^N$ as $N\rightarrow\infty$ and so, the $N-K$ basis vectors of $\cal{S}'$ are the $N-K$ linearly independent eigenvectors of $\Pi_{\mathbf{A}_{\tau}}^{\bot}$ corresponding to the eigenvalue $1$. Consequently, the main diagonal of $\mathbf{D}$ consists of $N-K$ $1$'s and $K$ $0$'s. Moreover we have: 
\begin{align}
&\lVert\Pi_{\mathbf{A}_{\tau}}^{\bot}\mathbf{y}\rVert ^2=\lVert\Pi_{\mathbf{A}_{\tau}}^{\bot}\mathbf{n}\rVert ^2=\lVert \mathbf{U}_{\tau}\mathbf{D}\mathbf{U}_{\tau}^T\mathbf{n}\rVert ^2=\nonumber \\ 
&\mathbf{n}^T\mathbf{U}_{\tau}\mathbf{D}\mathbf{U}_{\tau}^T\mathbf{U}_{\tau}\mathbf{D}\mathbf{U}_{\tau}^T\mathbf{n}=\nonumber \\ 
&(\mathbf{D}\mathbf{n}')^T\mathbf{D}\mathbf{n}'=\lVert \mathbf{D}\mathbf{n}'\rVert^2~,
\end{align} 
in which $\mathbf{n}'=\mathbf{U}_\tau^T\mathbf{n}$ is a white Gaussian random vector (according to the fact that $\mathbf{U}_\tau^T$ is just a deterministic rotation transform), i.e., $\mathbf{n}'\sim N(0,\sigma_n^2 \mathbf{I})$. Without loss of generality, we can assume that the $N-K$ first elements of $\mathbf{D}$ are $1$. So we can say that:
\begin{equation}
\varphi_1=\lVert \mathbf{D}\mathbf{n}'\rVert^2=\lvert n'_1\rvert^2+\lvert n'_2\rvert^2\dots+\lvert n'_{N-K}\rvert^2~.
\end{equation} 
Since $\varphi_1$ is the sum of squares of $N-K$ independent Gaussian random variables with mean $0$ and variance $\sigma_n^2$, it is a $\chi^2$ random variable of order $N-K$ with parameter $\sigma_n^2$. i.e.,
\begin{equation}
\mathbb{E}\{\varphi_1\}=(N-K)\sigma_n^2~~~~\textrm{var}\{\varphi_1\}=2(N-K)\sigma_n^4\nonumber~.
\end{equation}
This $\chi^2$ random variable has a moment generating function $\Phi_{\varphi_1}(s)$, which is defined by $\Phi_{\varphi_1}(s)\triangleq\mathbb{E}\{e^{\varphi_1 s}\}$, and for every $s$ satisfying the condition $1-2s\sigma_n^2>0$ can be expressed as~\cite{papoulis1965probability}:
\begin{equation}
\label{eqmgf1}
\Phi_{\varphi_1}(s)=\frac{1}{(1-2s\sigma_n^2)^{\frac{N-K}{2}}}~.
\end{equation}
We can rewrite the probability in (\ref{th1}) as follows:
\begin{equation}
\begin{split}
&\mathbb{P} \{\lvert \frac{1}{N}\lVert\Pi_{\mathbf{A}_{\tau}}^{\bot}\mathbf{y}\rVert ^2-\frac{N-K}{N}\sigma_{n}^2\rvert>\epsilon  \}=\\
&\mathbb{P} \{\lvert \varphi_1-(N-K)\sigma_n^2\rvert>N\epsilon  \} \leq\\
\label{eq3}
&\mathbb{P} \{ \varphi_1>N\epsilon+(N-K)\sigma_n^2  \} +\mathbb{P} \{ \varphi_1<-N\epsilon+(N-K)\sigma_n^2\}~.
\end{split}
\end{equation}
So, by using Chernof bounds on the tail probability~\cite{hagerup1990guided}, i.e.:
\begin{align}
\label{eq4}&\forall \nu >0 :\mathbb{P}\{\varphi_1 >\delta\}\leq e^{-\nu\delta}\Phi_{\varphi_1}(\nu)\\
\label{eq5}&\forall \nu <0 :\mathbb{P}\{\varphi_1 <\delta\}\leq e^{-\nu\delta}\Phi_{\varphi_1}(\nu)~,
\end{align}
we can bound the probabilities in (\ref{eq3}). By applying (\ref{eq4}) and (\ref{eqmgf1}), and also considering the constraints needed for these equations, we have:
\begin{align}
&\forall~0<\nu<\frac{1}{2\sigma_n^2} :\mathbb{P} \{ \varphi_1>N\epsilon+(N-K)\sigma_n^2  \}\leq\ \nonumber\\
&\frac{1}{(1-2\nu\sigma_n^2)^{\frac{N-K}{2}}}\exp\{-\nu(N\epsilon+(N-K)\sigma_n^2)\}\triangleq f(\nu)~.
\end{align}
By taking the derivative of $f(\nu)$ and finding its minimum in order to obtain the tightest bound, we find that this minimum occurs at $\nu^*=\frac{1}{2\sigma_n^2}\frac{N\epsilon}{N\epsilon+(N-K)\sigma_n^2}$. Moreover, it is easy to check that $\nu^*$ satisfies the constraints imposed by (\ref{eq4}) and (\ref{eqmgf1}), i.e. $\nu^*>0$ and $1-2\sigma_n^2\nu^*=\frac{(N-K)\sigma_n^2}{N\epsilon+(N-K)\sigma_n^2}>0$. Hence we have:
\begin{align}
&\mathbb{P} \{ \varphi_1>N\epsilon+(N-K)\sigma_n^2  \}\leq\ f(\nu^*)=\nonumber\\
&\Big\{\frac{N\epsilon+(N-K)\sigma_n^2}{(N-K)\sigma_n^2}\Big\}^{\frac{N-K}{2}}\exp\{-\frac{N\epsilon}{2\sigma_n^2}\}=\nonumber\\
&\exp\Big\{\frac{(N-K)}{2}\ln(1+\frac{N\epsilon}{(N-K)\sigma_n^2})-\frac{N\epsilon}{2\sigma_n^2}\Big\}=\nonumber\\
\label{eq6}&\exp\Big\{-\frac{(N-K)}{2}\Big(-\ln(1+\frac{\epsilon}{{{\sigma'}_n^{2}}})+\frac{\epsilon}{{\sigma'}_n^{2}}\Big)\Big\}~,
\end{align} 
in which ${\sigma'}_n^{2}=\frac{(N-K)}{N}\sigma_n^2$. Using the inequality $\ln(1+\frac{\epsilon}{{{\sigma'}_n^{2}}})\leq \frac{\epsilon}{{\sigma'}_n^{2}}$, we can say that the bound in (\ref{eq6}) decreases exponentially to $0$ as $N\rightarrow\infty$. Similarly, using (\ref{eq5}) and following the same approach as in the proof of (\ref{eq6}), we can bound $\mathbb{P} \{ \varphi_1<-N\epsilon+(N-K)\sigma_n^2\}$ and so we will have:
\begin{align}
&\mathbb{P} \{ \varphi_1<-N\epsilon+(N-K)\sigma_n^2\}\leq\nonumber\\
\label{eq7}&\exp\Big\{-\frac{(N-K)}{2}\Big(-\ln(1-\frac{\epsilon}{{{\sigma'}_n^{2}}})-\frac{\epsilon}{{\sigma'}_n^{2}}\Big)\Big\}~.
\end{align}  
Using the inequality  $\ln(1-\frac{\epsilon}{{{\sigma'}_n^{2}}})\leq -\frac{\epsilon}{{\sigma'}_n^{2}}$, it is seen that the bound in (\ref{eq7}) approaches $0$ exponentially as $N\rightarrow\infty$. Consequently the probability in (\ref{th1}) will tend at least exponentially to $0$, and so the proof is complete.  
%if we define $\varphi_2=\frac{\varphi_1}{N}$, we will have:
%\begin{equation}
%\mathbb{E}\{\varphi_2\}=\frac{(N-K)}{N}\sigma_n^2~~~~\textrm{var}\{\varphi_2\}=2\frac{(N-K)}{N^2}\sigma_n^4\nonumber
%\end{equation}
%Using the Chebyshev inequality, we can bound the probability of part (1) as follows:
%\begin{align}
%&\mathbb{P} \{\lvert \frac{1}{N}\lVert\Pi_{\mathbf{A}_{\tau}}^{\bot}\mathbf{y}\rVert ^2-\frac{N-K}{N}\sigma_{n}^2\rvert>\epsilon  \}=\mathbb{P} \{\lvert \varphi_2-\mathbb{E}\{\varphi_2\}\rvert>\epsilon  \}\leq\nonumber\\
%&\frac{\textrm{var}\{\varphi_2\}}{\epsilon^2}=\frac{2(N-K)}{N^2}\frac{\sigma_n^4}{\epsilon^2}\rightarrow 0 ~~~\text{as} ~~~N\rightarrow\infty
%\end{align}
%which will complete the proof of part (1).
\end{proof}

\begin{proof}[\textbf{Proof of (\ref{th2})}]
Similar to the previous part we have:
\begin{align}
&\Pi_{\mathbf{A}_{\xi}}^{\bot}\mathbf{y}=\Pi_{\AT}^{\bot}(\ATa\mathbf{s}_{\tau}+\,\mathbf{n})=\nonumber\\
&\Pi_{\AT}^{\bot}(\sum_{i\in \tau\cap\xi}{s_i\mathbf{a}_i}+\sum_{i\in \tau\backslash\xi}{s_i\mathbf{a}_i}+\mathbf{n})=\Pi_{\AT}^{\bot}(\sum_{i\in \tau\backslash\xi}{s_i\mathbf{a}_i}+\mathbf{n})\nonumber~.
\end{align}
In the same way, we can decompose $\Pi_{\AT}^{\bot}=\mathbf{U}_{\xi}\mathbf{D}\mathbf{U}_{\xi}^T$, in which $\mathbf{D}$ is similar to the one in the previous part and $\mathbf{U}_{\xi}$ is a unitary matrix. Then we have:
\begin{align}
&\lVert\Pi_{\mathbf{A}_{\xi}}^{\bot}\mathbf{y}\rVert ^2=\lVert \mathbf{U}_{\xi}\mathbf{D}\mathbf{U}_{\xi}^T{(\sum_{i\in \tau\backslash\xi}{s_i\mathbf{a}_i}+\mathbf{n})}\rVert ^2=\lVert \mathbf{D}\mathbf{n}''\rVert ^2\nonumber~,
\end{align}
in which, $\mathbf{n}''=\mathbf{U}_{\xi}^T\mathbf{n}+\sum_{i\in \tau\backslash\xi}{s_i \mathbf{U}_{\xi}^T \mathbf{a}_i}$ is a Gaussian random vector with mean $\bar{\mathbf{n}}''=\mathbb{E}\{\mathbf{n}''\}=\sum_{i\in \tau\backslash\xi}{s_i \mathbf{U}_{\xi}^T \mathbf{a}_i}$ and auto-covariance matrix  $\mathbb{E}\{(\mathbf{n}''-\bar{\mathbf{n}}'')({\mathbf{n}''}-\bar{\mathbf{n}}'')^T\}=C_{\mathbf{n}''}=\sigma_{n}^2\mathbf{I}$, which are results of the fact that $\mathbf{A}$ is deterministic. It is important to note that the remaining proof of this part of Lemma 3.3 in~\cite{akcakaya711shannon} (which is so similar to our proposed lemma) is based on the Gaussian assumption on $\mathbf{A}$, in addition to the assumption that this matrix is random in the domain of noise\footnote{This approach is very common in the framework of information theory, when one tries to show the achievability of a rate in a channel~\cite{cover2006elements}.}; nevertheless, our proof is free from such assumptions while we assume that the measurement matrix is deterministic. As a result, this assumption will help us to generalize our results for other types of randomly generated measurement matrices that are common in the compressed sensing area, as will be shown later in this correspondence. 

To continue our proof, without loss of generality we can assume that the first $N-K$ elements of the main diagonal of $\mathbf{D}$ are $1$ and so:
\begin{equation}
\varphi_2=\lVert \mathbf{D}\mathbf{n}''\rVert^2=\lvert n''_1\rvert^2+\lvert n''_2\rvert^2\dots+\lvert n''_{N-K}\rvert^2~,
\end{equation}
in which, $n''_i\sim N(m_i,\sigma_n^2)$ for every $1\leq i\leq N-K$. In addition to these, we have:
\begin{equation}
\label{eq1}
\sum_{i=1}^{N-K}{m_i^2}=\lVert\sum_{i\in\tau\backslash\xi}{(\mathbf{U}_{\xi}^T\mathbf{a}_i)_{N-K}s_i} \rVert ^2=\sum_{i\in\tau\backslash\xi}{\sum_{j\in\tau\backslash\xi}{s_i s_j \mathbf{a'}_i^T\mathbf{a'}_j}}~,
\end{equation}
in which, the $(\cdot)_{N-K}$ operator denotes a sub-vector of the first $N-K$ elements and $\mathbf{a'}_i =(\mathbf{U}_{\xi}^T\mathbf{a}_i)_{N-K}$. For the sake of simplicity of notations, we define $\gamma^2=\frac{1}{N}\sum_{i\in\tau\backslash\xi}{\sum_{j\in\tau\backslash\xi}{s_i s_j \mathbf{a'}_i^T\mathbf{a'}_j}}$. Now, the sum of the squares of $N-K$ independent Gaussian random variables $n''_i$, each having mean $m_i$, is a \textit{non-central} $\chi^2$ random variable of order $N-K$ with parameters $\sigma_n^2$ and $\sum_{i=1}^{N-K}{m_i^2}$. So we have:
\begin{equation}
\mathbb{E}\{\varphi_2\}=(N-K)\sigma_n^2+N\gamma^2~,~~~\textrm{var}\{\varphi_2\}=2(N-K)\sigma_n^4+4\sigma_n^2N^2\gamma^4\nonumber~.
\end{equation}
Additionally, this $\chi^2$ random variable has a moment generating function $\Phi_{\varphi_2}(s)$, which is defined by $\Phi_{\varphi_2}(s)\triangleq\mathbb{E}\{e^{\varphi_2 s}\}$, and for every $s$ satisfying $1-2s\sigma_n^2>0$ can be expressed as\cite{papoulis1965probability}: 
\begin{equation}
\label{eq77}
\Phi_{\varphi_2}(s)=\frac{1}{(1-2s\sigma_n^2)^{\frac{N-K}{2}}}\exp\Big(\frac{s\sum_{i=1}^{N-K}{m_i^2}}{1-2s\sigma_n^2} \Big)~.
\end{equation}
By \textit{centralizing} the probability in (\ref{th2}) with respect to the mean of $\varphi_2$, we can rewrite the probability in (\ref{th2}) as:
\begin{align}
\mathbb{P}\{\lvert &\frac{1}{N}\lVert\Pi_{\mathbf{A}_{\xi}}^{\bot}\mathbf{y}\rVert ^2-\frac{N-K}{N}\sigma_{n}^2\rvert<\epsilon  \}=\nonumber\\
&\mathbb{P}\{\lvert \varphi_2-(N-K)\sigma_n^2\rvert < N\epsilon  \}\leq\mathbb{P}\{\varphi_2-(N-K)\sigma_n^2 < N\epsilon  \}=\nonumber\\
&\mathbb{P}\{\varphi_2-(N-K)\sigma_n^2 < N(\gamma^2-\bar{\epsilon}) \}=\nonumber\\ 
\label{eq8}&\mathbb{P}\{\varphi_2-(N-K)\sigma_n^2-N\gamma^2 <-N\bar{\epsilon}\},
\end{align} 
in which $\bar{\epsilon}=\gamma^2-\epsilon>0$ (we assume that $\epsilon$ is small enough so that $\epsilon<\gamma^2$). Similar to the proof of (\ref{th1}), we will use Chernof bounds stated in (\ref{eq4}) and (\ref{eq5}) to bound the probability in (\ref{eq8}). More accurately, by the use of (\ref{eq5}) we get:

\begin{align}
\label{eq10}
&\forall \nu < 0:~\mathbb{P}\{\varphi_2<(N-K)\sigma_n^2+N\gamma^2-N\bar{\epsilon}\}\leq\nonumber\\
&\exp\Big(-\nu[(N-K)\sigma_n^2+N\gamma^2-N\bar{\epsilon}]\Big)\Phi_{\varphi_2}(\nu)\triangleq g(\nu)~.
\end{align} 
By plugging in the value of $\Phi_{\varphi_2}(\nu)$ from (\ref{eq77}) for every $\nu$ satisfying $1-2\nu\sigma_n^2>0$, $g(\nu)$  is equal to:
\begin{align}
\label{eqmgf2}
&g(\nu)=\frac{1}{(1-2\nu\sigma_n^2)^{\frac{N-K}{2}}}\times\nonumber\\ 
&\exp\Big(\frac{\nu\sum_{i=1}^{N-K}{m_i^2}}{1-2\nu\sigma_n^2} -\nu[(N-K)\sigma_n^2+N\gamma^2-N\bar{\epsilon}]\Big)~.
\end{align}
As shown in the Appendix (Lemma~\ref{lem1}), by taking the derivative of $g(\nu)$ with respect to $\nu$, one can see that this function will reach its minimum value at $\nu^*$, calculated as following:
\begin{align}
\label{eq11}
\nu^*=\frac{2\gamma^2-2\bar{\epsilon}+{\sigma'}_n^{2}-\sqrt{({\sigma'}_n^{2}+2\gamma^2)^2-4\gamma^2\bar{\epsilon}}}{4\sigma_n^{2}(\gamma^2-\bar{\epsilon}+{\sigma'}_n^{2})}~,
\end{align}
in which ${\sigma'}_n^{2}=\frac{N-K}{N}\sigma_n^{2}$. Moreover, this $\nu^*$ is negative (and hence satisfies the constraint $1-2\nu^*\sigma_n^2>0$), as stated by Lemma~A.1.
By plugging (\ref{eq11}) and the expressions obtained in the Appendix for $1-2\nu^*\sigma_n^2$, $\frac{1}{1-2\nu^*\sigma_n^2}$ and $\frac{\gamma^2\nu^*}{1-2\nu^*\sigma_n^2}$ in (\ref{eq10}), we will have the tightest Chernof bound for the probability in (\ref{eq10}) as:
\begin{align}
\label{eq14}
&g(\nu^*)=\Big[\frac{\sqrt{({\sigma'}_n^{2}+2\gamma^2)^2-4\gamma^2\bar{\epsilon}}-{\sigma'}_n^{2}}{2\gamma^2}\}\Big]^{\frac{N-K}{2}}\times\nonumber\\
&\exp\Big\{-N\big[ \frac{{\sigma'}_n^{2}+2\gamma^2-\bar{\epsilon}-\sqrt{({\sigma'}_n^{2}+2\gamma^2)^2-4\gamma^2\bar{\epsilon}}}{2\sigma_n^{2}}  \big]        \Big\}~.
\end{align}
After some manipulations, this bound can be re-written as:
\begin{align}
\label{eq12}
&\exp\big\{-\frac{N}{2}\Big[ \frac{{\sigma'}_n^{2}+2\gamma^2-\bar{\epsilon}-({\sigma'}_n^{2}+2\gamma^2)\sqrt{1-\frac{4\gamma^2\bar{\epsilon}}{({\sigma'}_n^{2}+2\gamma^2)^2}}}{\sigma_n^{2}} \nonumber\\ 
&-\frac{N-K}{N}\ln\big( \frac{({\sigma'}_n^{2}+2\gamma^2)\sqrt{1-\frac{4\gamma^2\bar{\epsilon}}{({\sigma'}_n^{2}+2\gamma^2)^2}}-{\sigma'}_n^{2}}{2\gamma^2} \big)\Big]        \big\}~.
\end{align}

 Before proceeding any further, we will introduce the following lemma:
\begin{lemma}
\label{lem2}
for any $x\in\mathbb{R}$ and $0\leq x\leq 1	$ we have:
\begin{itemize}
\item $\sqrt{1-x}~~\leq 1-\frac{1}{2}x$
\item $\ln(1-x)\leq -x-\frac{1}{2}x^2$
\end{itemize}
\end{lemma}   
the proof is elementary and is left to the reader.

It is important to note that $\frac{4\gamma^2\bar{\epsilon}}{({\sigma'}_n^{2}+2\gamma^2)^2}=\frac{4\gamma^4-4\gamma^2\epsilon}{4\gamma^4+{\sigma'}_n^4+2\gamma^2{\sigma'}_n^2}$, and so we have that $0\leq \frac{4\gamma^2\bar{\epsilon}}{({\sigma'}_n^{2}+2\gamma^2)^2}\leq 1$. Hence, by using the first part of Lemma~(\ref{lem2}) and some further manipulations, we can bound (\ref{eq14}) as:
\begin{align}
\label{eq13}
&g(\nu^*)\leq\exp\big\{-\frac{N}{2}\Big[ \frac{{\sigma'}_n^{2}+2\gamma^2-\bar{\epsilon}-({\sigma'}_n^{2}+2\gamma^2)+\frac{2\gamma^2\bar{\epsilon}}{({\sigma'}_n^{2}+2\gamma^2)}}{\sigma_n^{2}} \nonumber\\ 
&-\frac{N-K}{N}\ln\big( \frac{({\sigma'}_n^{2}+2\gamma^2)-\frac{2\gamma^2\bar{\epsilon}}{({\sigma'}_n^{2}+2\gamma^2)}-{\sigma'}_n^{2}}{2\gamma^2} \big)\Big]        \big\}~.
\end{align}
After simplifying (\ref{eq13}) we have:
\begin{align}
\label{eq15}
&g(\nu^*)\leq\exp\big\{-\frac{N}{2}\Big[ -\frac{{\sigma'}_n^{2}\bar{\epsilon}}{\sigma_n^2({\sigma'}_n^{2}+2\gamma^2)} \nonumber \\
&~~~~~~-\frac{N-K}{N}\ln\big( 1-\frac{\bar{\epsilon}}{({\sigma'}_n^{2}+2\gamma^2)} \big)\Big]        \big\}=\nonumber\\
&\exp\big\{-\frac{N-K}{2}\Big[ -\frac{\bar{\epsilon}}{({\sigma'}_n^{2}+2\gamma^2)} 
-\ln\big( 1-\frac{\bar{\epsilon}}{({\sigma'}_n^{2}+2\gamma^2)} \big)\Big]        \big\}~.
\end{align}
It is also important to note that as $\frac{\bar{\epsilon}}{({\sigma'}_n^{2}+2\gamma^2)}=\frac{\gamma^2-{\epsilon}}{({\sigma'}_n^{2}+2\gamma^2)}$, we have $0 \leq \frac{\bar{\epsilon}}{({\sigma'}_n^{2}+2\gamma^2)}\leq 1$. Now, by applying the second part of Lemma~\ref{lem2} we can make an upper bound for (\ref{eq15}) as the following:
\begin{equation}
\label{eq16}
g(\nu^*)\leq\exp\big\{-\frac{N-K}{4}\Big[\frac{\bar{\epsilon}}{({\sigma'}_n^{2}+2\gamma^2)} 
\Big]^2       \big\}~.
\end{equation}
 Therefore, according to (\ref{eq16}), (\ref{eq10}) and (\ref{eq8}), we have finally come to the following result:
\begin{align}
&\mathbb{P}\{\lvert \frac{1}{N}\lVert\Pi_{\mathbf{A}_{\xi}}^{\bot}\mathbf{y}\rVert ^2-\frac{N-K}{N}\sigma_{n}^2\rvert<\epsilon  \}\leq\nonumber\\
&\exp\big\{-\frac{N-K}{4}\Big[\frac{\bar{\epsilon}}{({\sigma'}_n^{2}+2\gamma^2)} 
\Big]^2       \big\}=\nonumber\\
&\exp\big\{-\frac{N-K}{4}\Big[\frac{\frac{1}{N}\sum_{i\in\tau\backslash\xi}{\sum_{j\in\tau\backslash\xi}{s_i s_j \mathbf{a'}_i^T\mathbf{a'}_j}}-\epsilon}{\frac{2}{N}\sum_{i\in\tau\backslash\xi}{\sum_{j\in\tau\backslash\xi}{s_i s_j \mathbf{a'}_i^T\mathbf{a'}_j}}+{\sigma'}_n^{2}} 
\Big]^2       \big\}~,
\end{align}
and this will complete the proof of (\ref{th2}).
\end{proof}  
 \end{theorem}

It is important to note that if we see the proof of Theorem 2.1, then we will conclude that this theorem holds asymptotically in probability, i.e. if you test the validity of Theorem 2.1 for infinite numbers of randomly generated $\mathbf{A}$, then this theorem may not be valid for just finite numbers of $\mathbf{A}$. Moreover, as $N\to \infty$ the size of this finite set will tend to zero. Accordingly, one can say that as $N\to\infty$, Theorem 2.1 may not be valid for just asymptotic \textit{zero} number of randomly generated $\mathbf{A}$, or simply it is asymptotically valid. However, as we will see later in Section \Rmnum{3}, we want to consider the achievability of Cram\'{e}r-Rao bound in asymptotic case, and so this asymptotic validation should be enough. 

In addition to what has been stated in Theorem 2.1, when the size of the problem tends to infinity and $\mathbf{A}$ satisfies the introduced concentration of measures inequality depicted in (\ref{con2:eq}) (for instance, its elements are drawn i.i.d from $N(0,1)$ or distributions such as the ones introduced in (\ref{dis1}) and (\ref{dis2})), one may find an equivalent bound using the following lemma:
\begin{lemma}
\label{lem3}
If the elements of $\mathbf{A}$ are randomly and independently generated according to a distribution that satisfies (\ref{con2:eq}), then we have: 
\begin{equation}
\sum_{i\in\tau\backslash\xi}{\sum_{j\in\tau \backslash \xi}{s_i s_j \mathbf{a'}_i^T\mathbf{a'}_j}}\rightarrow (N-K)\sum_{i\in\tau\backslash\xi}{\lvert s_i\rvert ^2}~~,
\end{equation} 
in which $\tau$, $\xi$, $\mathbf{s}$ and $\mathbf{a'}_i$ are defined as in Theorem~2.1 .
\end{lemma}
\begin{proof}
Suppose that $\mathbf{x_1}$ and $\mathbf{x_2}$ are two arbitrary fixed vectors in $\mathbb{R}^M$. Then for every $\epsilon \in (0,1)$, the following inequalities hold with a probability that tends exponentially to 1 as $N$ tends to $\infty$:
\begin{align}
&\label{x1}(1-\epsilon)N\lVert \mathbf{x}_1\rVert^2\leq\lVert\mathbf{A}\mathbf{x}_1\rVert^2\leq(1+\epsilon)N\lVert \mathbf{x}_1\rVert^2~, \\
&\label{x2}(1-\epsilon)N\lVert \mathbf{x}_2\rVert^2\leq\lVert\mathbf{A}\mathbf{x}_2\rVert^2\leq(1+\epsilon)N\lVert \mathbf{x}_2\rVert^2~, \\
&\label{x3}(1-\epsilon)N\lVert \mathbf{x}_1-\mathbf{x}_2\rVert^2\leq\lVert\mathbf{A}(\mathbf{x}_1-\mathbf{x}_2)\rVert^2\leq(1+\epsilon)N\lVert \mathbf{x_1}-\mathbf{x_2}\rVert^2~.
\end{align}
Using (\ref{x1}), (\ref{x2}) and (\ref{x3}), it is straightforward to show that:
\begin{align}
&(1+\epsilon)N\mathbf{x}_1^T\mathbf{x}_2-\epsilon N(\lVert \mathbf{x}_1\rVert^2+\lVert \mathbf{x}_2\rVert^2)\leq\nonumber\\
&(\mathbf{A}\mathbf{x}_1)^T(\mathbf{A}\mathbf{x}_2)\leq\nonumber\\
&(1-\epsilon)N\mathbf{x}_1^T\mathbf{x}_2+\epsilon N(\lVert \mathbf{x}_1\rVert^2+\lVert \mathbf{x}_2\rVert^2)~.
\end{align}
By setting $\mathbf{x_1}$ and $\mathbf{x_2}$ with $1$'s in their $i$-th and $j$-th elements respectively and $0$'s in their other elements, for $i\neq j$ we will have:
\begin{equation}
-2\epsilon\leq\frac{1}{N}\mathbf{a}_i^T\mathbf{a}_j\leq2\epsilon~,
\end{equation} 
and in the case of $i=j$, we will have:
\begin{equation}
1-\epsilon\leq\frac{1}{N}\mathbf{a}_i^T\mathbf{a}_i\leq 1+\epsilon~.
\end{equation} 
These events hold valid with a probability that tends exponentially to 1 as $N$ tends to $\infty$ for a fix value of $i$ and $j$. By applying the union bound on all $\binom{M}{2}=\frac{M(M-1)}{2}$ choices for $i$ and $j$, if $\epsilon \rightarrow 0$ then the  following equation holds for every $i$ and $j$, with a probability that still tends to 1 as $N$ increases: 
\begin{equation}
\label{eq2}
\frac{1}{N}\mathbf{a}_i^T\mathbf{a}_j\rightarrow 0~~\text{if}~~ i\neq j~,~\frac{1}{N}\mathbf{a}_i^T\mathbf{a}_j\rightarrow 1 ~~\text{if}~~ i=j~.
\end{equation}

%Obviously, $\mathbf{\Phi}=\mathbf{U}_{\xi}^T\mathbf{A}$ also satisfies the concentration of measure inequality when $\mathbf{A}$ satisfies it, according to the fact that $\mathbf{U}_{\xi}$ is an unitary matrix. Though, by doing the same manipulations we will have a similar equation to (\ref{eq2}) for $\mathbf{a'}_i=(\mathbf{U}_{\xi}^T\mathbf{a}_i)_{(N-K)}$:
%\begin{equation}
%\label{eq23}
%\mathbf{a'}_i^T\mathbf{a'}_j\rightarrow 0~~\text{if}~~ i\neq j~,~\mathbf{a'}_i^T\mathbf{a'}_j\rightarrow \frac{N-K}{N} ~~\text{if}~~ i=j~.
%\end{equation}
%Substituting (\ref{eq23}) in (\ref{eq1}) will complete the proof. 

%It is well-known in the literature of compressed sensing that if the elements of $\mathbf{A}_{N\times M}$ are generated randomly from a $N(0,1)$ distribution, then for a large enough $N$, every $N\times N$ sub-matrix of $\mathbf{A}$ will be proportional to a unitary matrix with probability 1~\cite{davenport2006detection,baraniuk2008simple}, i.e. its columns will be orthogonal and $\lVert\mathbf{a}_i\rVert_2^2\rightarrow N$ as $N\rightarrow\infty$ with probability 1\footnote{this can easily be seen by applying the law of large numbers on the inner product summation of two identical or different columns of $\mathbf{A}$ and using the union bound to bound the probability. We omit the proof for the lack of space.}. 

Now, consider the matrix $(\mathbf{U}_\xi^T\mathbf{A})_{N-K}$. We want to show this $(N-K)\times M$ matrix will also satisfy the modified version of the concentration of measures inequality. In other words, we want to show that for every $\mathbf{x}\in \mathbb{R}^M$, the following equation holds with a probability that tends exponentially to $1$ as $N$ tends $\infty$:
\begin{equation}
\label{eqadd1}
(1-\epsilon)(N-K)\lVert \mathbf{x}\rVert^2\leq\lVert(\mathbf{U}_{\xi}^T\mathbf{A})_{N-K}\mathbf{x}\rVert^2\leq(1+\epsilon)(N-K)\lVert \mathbf{x}\rVert^2\enspace.
\end{equation}

To show this, we have:
\begin{equation}
\label{eqadd2}
\lVert(\mathbf{U}_{\xi}^T\mathbf{A})_{N-K}\mathbf{x}\rVert^2=\mathbf{x}^T\mathbf{A}^T(\mathbf{U}_{\xi}^T)_{N-K}^T(\mathbf{U}_{\xi}^T)_{N-K}\mathbf{A}\mathbf{x}\enspace.
\end{equation}
To simplify (\ref{eqadd2}), lets see how the matrix $\mathbf{U}_{\xi}$ is constructed. First, choose a set of indices in $\{1,2,\dots,M\}$ such as $\mathscr{L}\subset\{1,2,...M\}$, so that $\lvert\mathscr{L}
\rvert=N$, and also $\xi\subset \mathscr{L}$ and $\tau\subset \mathscr{L}$. Then, we choose $N$ columns of $\mathbf{A}_{\textrm{norm}}=\frac{1}{\sqrt{N}}\mathbf{A}$ corresponding to the indices in $\mathscr{L}$. Following (\ref{eq2}), we can say that the columns of $\mathbf{A}_{\textrm{norm}}$ corresponding to the indices in $\xi$ are an \textit{approximate} orthonormal basis for the span of columns of $\mathbf{A}_{\xi}$ with a probability that tends exponentially to $1$ as $N\to\infty$, and this approximation will become more accurate as $\epsilon$ is chosen smaller. Therefore, these columns can be considered asymptotically as approximations for the orthonormal eigenvectors of the symmetric matrix $\Pi_{\mathbf{A}_{\xi}}^{\bot}$ corresponding to zero eigenvalue, and again these approximations will become more accurate as $\epsilon$ is chosen smaller. Similarly, the columns of  $\mathbf{A}_{\textrm{norm}}$ corresponding to the indices in $\mathscr{L}\backslash\xi$ can be considered as an approximate orthonormal basis for the kernel space of $\mathbf{A}_{\xi}$ with a probability that tends exponentially to $1$, and so they are approximations for orthonormal eigenvectors of $\Pi_{\mathbf{A}_{\xi}}^{\bot}$ corresponding to the eigenvalue 1. Consequently, by the definition of $\mathbf{U}_{\xi}$ (i.e. its first $N-K$ columns are orthonormal eigenvectors of $\Pi_{\mathbf{A}_{\xi}}^{\bot}$ corresponding to eigenvalue 1 and the next $K$ column are orthonormal eigenvectors of $\Pi_{\mathbf{A}_{\xi}}^{\bot}$ corresponding to zero eigenvalue) and the approximate orthogonal property of the selected columns of $\mathbf{A}$ as $N\to\infty$ (equation (\ref{eq2})), and by doing some simple manipulations, we have:
\begin{equation}
\label{eqadd3}
(\mathbf{U}_{\xi}^T)_{N-K}^T(\mathbf{U}_{\xi}^T)_{N-K}\approx\frac{N-K}{N}\mathbf{I}_{M\times M}\enspace ,
\end{equation}
and this approximation will become more accurate as $\epsilon$ is chosen smaller. By substituting the approximation stated in (\ref{eqadd3}) with corresponding term in (\ref{eqadd2}) we have:
\begin{equation}
\label{eqadd4}
\lVert(\mathbf{U}_{\xi}^T\mathbf{A})_{N-K}\mathbf{x}\rVert^2\approx \frac{N-K}{N}\lVert\mathbf{A}\mathbf{x}\rVert^2\enspace,
\end{equation}
and again, this approximation will be more accurate with smaller $\epsilon$. So, following (\ref{con2:eq}), one can say that for small enough $\epsilon$ the equation (\ref{eqadd1}) holds with a probability that tends exponentially to $1$ as $N$ grows~\footnote{Note that for small enough $\epsilon$ we require large enough $N$ (following what has been stated in (\ref{con2:eq})), so that the concentration of measures inequality will be satisfied with high probability.}. Now, using (\ref{eqadd1}) and similar to what we have stated about the columns of $\mathbf{A}$, we can conclude that the columns of $(\mathbf{U}_{\xi}^T\mathbf{A})_{N-K}=[\mathbf{a'}_1,\mathbf{a'}_2,\dots,\mathbf{a'}_M]$ satisfy the following equation as $N\to\infty$: 

\begin{equation}
\label{eq3}
\frac{1}{(N-K)}\mathbf{a'}_i^T\mathbf{a'}_j\rightarrow 0~~\text{if}~~ i\neq j~,~\frac{1}{(N-K)}\mathbf{a'}_i^T\mathbf{a'}_j\rightarrow 1 ~~\text{if}~~ i=j~.
\end{equation} 
Substituting (\ref{eq3}) in (\ref{eq1}) will complete the proof. 
\end{proof} 

%It is worth mentioning that for measurement matrices that satisfy (\ref{con:eq}) after a normalization, such as the ones with i.i.d $N(0,1)$ entries or examples introduced in (\ref{dis1}) and (\ref{dis2}), equation (\ref{eq23}) will also be valid with a change in the value of $\mathbf{a'}_i^T\mathbf{a'}_j$ in the case of $i=j$. Indeed, we have the following equation for the mentioned measurement matrices:
%\begin{equation}
%\label{eq24}
%\mathbf{a'}_i^T\mathbf{a'}_j\rightarrow 0~~\text{if}~~ i\neq j~,~\mathbf{a'}_i^T\mathbf{a'}_j\rightarrow N-K ~~\text{if}~~ i=j~.
%\end{equation}

Now, using Lemma~\ref{lem3}, we can rewrite the bound in (\ref{th2}) after some manipulations as:

\begin{align}
&\mathbb{P}\{\lvert \frac{1}{N}\lVert\Pi_{\mathbf{A}_{\xi}}^{\bot}\mathbf{y}\rVert ^2-\frac{N-K}{N}\sigma_{n}^2\rvert<\epsilon  \}\leq\nonumber\\
\label{eq18}&\exp\big\{-\frac{N-K}{4}\Big[\frac{\sum_{i\in\tau\backslash\xi}{\lvert s_i\rvert ^2}-\epsilon'}{2\sum_{i\in\tau\backslash\xi}{\lvert s_i\rvert ^2}+{\sigma}_n^{2}} 
\Big]^2       \big\}~,
\end{align}
in which $\epsilon'=\frac{N}{N-K}\epsilon$. Interestingly, the asymptotic bound obtained in (\ref{eq18}) is very similar to the bound obtained in Lemma 3.3 of \cite{akcakaya711shannon}. In fact, the bound obtained in~\cite{akcakaya711shannon} is as the following:
\begin{align}
&\mathbb{P}\{\lvert \frac{1}{N}\lVert\Pi_{\mathbf{A}_{\xi}}^{\bot}\mathbf{y}\rVert ^2-\frac{N-K}{N}\sigma_{n}^2\rvert<\epsilon  \}\leq\nonumber\\
&\exp\big\{-\frac{N-K}{4}\Big[\frac{\sum_{i\in\tau\backslash\xi}{\lvert s_i\rvert ^2}-\epsilon'}{\sum_{i\in\tau\backslash\xi}{\lvert s_i\rvert ^2}+{\sigma}_n^{2}} 
\Big]^2       \big\}~.
\end{align}
 Although these bounds are not identical, but they are very similar.

\section{Overview of  Cram\'{e}r-Rao lower bound and the \\Jointly Typical Estimator}
In this section, we will discuss the problem of estimating $\mathbf{s}$ from noisy observations. The estimation process has two phases. In the first phase, the estimator will detect $\tau=\textrm{supp}(\mathbf{s})=\{i_1,i_2,\dots i_K\}$ which is the location of the taps. The second phase includes estimating $\mathbf{s}_{\tau}=[s_{i_1},s_{i_2},\dots,s_{i_K}]^T$ which is the value of the taps. In our discussion, we are going to survey the Cram\'{e}r-Rao lower bound of the estimation problem. By using the idea of two-phase estimation, we consider two special kinds of estimation process in this work. In the first case, the estimator has a complete prior knowledge of $\tau$, i.e. a genie has aided us with $\tau$. In the second case, we have no prior knowledge of $\tau$ except for its cardinality, $K$, which shows the level of sparsity. We will then derive that these two bounds are asymptotically equal to each other and are achievable by typical estimation, as shown in \cite{babadi2009asymptotic} although the main theorem used in \cite{babadi2009asymptotic} has been changed.

The model in (\ref{eq:ncs}) can be rewritten as:
\begin{equation}
\label{eq:ncs2}
\mathbf{y}={\Ab}\mathbf{s}+\mathbf{n}=\Ab_{\tau}\mathbf{s}_{\tau}+\mathbf{n}~.
\end{equation}

Now, if the estimator knows $\tau$ and wants to estimate $\mathbf{s}_{\tau}$ from $\mathbf{y}$ and $\tau$, then the Cram\'{e}r-Rao bound of the estimation can be computed using the following theorem, stated in~\cite{babadi2009asymptotic,haupt2010toeplitz}:

\begin{theorem}[Cram\'{e}r-Rao bound of genie aided estimation]
Considering the model depicted in (\ref{eq:ncs}) and estimators of the form $f(\mathbf{y},\tau)=\mathbf{\hat{s}}_{\tau}$, the Fisher information matrix of the GAE, which is defined as:
\begin{equation}
J_{\textrm{GAE}}=\mathbb{E}\{\Big[\frac{\partial}{\partial \mathbf{s}_{\tau}}\log{\mathbb{P}(\mathbf{y}|\mathbf{s}_{\tau})}\Big]\Big[\frac{\partial}{\partial \mathbf{s}_{\tau}}\log{\mathbb{P}(\mathbf{y}|\mathbf{s}_{\tau})}\Big]^T \}~,
\end{equation}
is equal to:
\begin{equation}
J_{\textrm{GAE}}=\frac{1}{\sigma_n^2}\mathbf{A}_{\tau}^T\mathbf{A}_{\tau}~,
\end{equation}
and so we have the following Cram\'{e}r-Rao bound\footnote{The equation $\mathbf{A}\geq\mathbf{B}$ means that $\mathbf{A}-\mathbf{B}$ is non-negative definite.} for the estimator $\mathbf{\hat{s}}_{\tau}=f(\mathbf{y},\tau)$:
\begin{align}
&\mathbb{E}\{(\mathbf{s}_{\tau}-\mathbf{\hat{s}}_{\tau})(\mathbf{s}_{\tau}-\mathbf{\hat{s}}_{\tau})^T\}\geq J^{-1}=\sigma_n^2(\mathbf{A}_{\tau}^T\mathbf{A}_{\tau})^{-1}\\
&\label{crb1}\mathbb{E}\{\lVert\mathbf{s}_{\tau}-\mathbf{\hat{s}}_{\tau}\rVert^2\}\geq \sigma_n^2\textrm{Trace}[(\mathbf{A}_{\tau}^T\mathbf{A}_{\tau})^{-1}]=\textrm{CRB-S}~.
\end{align}
\end{theorem}

\begin{proof}
The proof is given in \cite{babadi2009asymptotic} and \cite{haupt2010toeplitz}.
\end{proof}
In a GAE, by using a simple least square estimator for the model of (\ref{eq:ncs2}) we can achieve the Cram\'{e}r-Rao bound mentioned in (\ref{crb1}), i.e. this estimator is \textit{efficient}. In a more mathematical way, we have the following theorem:
\begin{theorem}[Structural Least Square Estimator (SLSE)]
Consider the following genie aided estimator 
\begin{equation*}
\mathbf{\hat{s}}_{\tau}=f(\mathbf{y},\tau)=\textrm{argmin}~~\lVert\mathbf{y}-\mathbf{A}_{\tau}\mathbf{x}_\tau \rVert=(\mathbf{A}_{\tau}^T\mathbf{A}_{\tau})^{-1}\mathbf{A}_{\tau}^T\mathbf{y}~,
\end{equation*}
then we have:
\begin{equation}
\mathbb{E}\{\lVert\mathbf{s}_{\tau}-\mathbf{\hat{s}}_{\tau}\rVert^2\}= \sigma_n^2\textrm{Trace}[(\mathbf{A}_{\tau}^T\mathbf{A}_{\tau})^{-1}]
\end{equation}
\begin{proof}
%we can write:
%\begin{align}
%&\mathbb{E}\{\lVert\mathbf{s}_{\tau}-\mathbf{\hat{s}}_{\tau}\rVert^2\}=
%\mathbb{E}\{\lVert\mathbf{s}_{\tau}-(\mathbf{A}_{\tau}^T\mathbf{A}_{\tau})^{-1}\mathbf{A}_{\tau}^T
%(\mathbf{A}_{\tau}\mathbf{s}_{\tau}+\mathbf{n})\rVert^2\}=\nonumber\\
%&\mathbb{E}\{\mathbf{n}^T\mathbf{A}_{\tau}(\mathbf{A}_{\tau}^T\mathbf{A}_{\tau})^{-1}(\mathbf{A}_{\tau}^T\mathbf{A}_{\tau})^{-1}\mathbf{A}_{\tau}^T\mathbf{n}\}=\nonumber\\
%&\textrm{Trace}\Big\{\mathbb{E}\{\mathbf{n}^T\mathbf{A}_{\tau}(\mathbf{A}_{\tau}^T\mathbf{A}_{\tau})^{-1}(\mathbf{A}_{\tau}^T\mathbf{A}_{\tau})^{-1}\mathbf{A}_{\tau}^T\mathbf{n}\}\Big\}=\nonumber\\
%&\mathbb{E}\Big\{\textrm{Trace}\{\mathbf{n}^T\mathbf{A}_{\tau}(\mathbf{A}_{\tau}^T\mathbf{A}_{\tau})^{-1}(\mathbf{A}_{\tau}^T\mathbf{A}_{\tau})^{-1}\mathbf{A}_{\tau}^T\mathbf{n}\}\Big\}=\nonumber\\
%&\mathbb{E}\Big\{\textrm{Trace}\{(\mathbf{A}_{\tau}^T\mathbf{A}_{\tau})^{-1}\mathbf{A}_{\tau}^T\mathbf{n}\mathbf{n}^T\mathbf{A}_{\tau}(\mathbf{A}_{\tau}^T\mathbf{A}_{\tau})^{-1}\}\Big\}=\nonumber\\
%&\sigma_n^2\textrm{Trace}\{(\mathbf{A}_{\tau}^T\mathbf{A}_{\tau})^{-1}\}~,
%\end{align}
The proof is similar to the proof of achievability of the Cram\'{e}r-Rao bound by the least square estimator where noise is Gaussian~\cite{kay1993fundamentals} and is omitted due to the lack of space. 
\end{proof}
\end{theorem}
When considering the asymptotic case in the estimation process, one may use the equivalent limit of the bound in (\ref{crb1}) using the following lemma:
\begin{lemma}
If the elements of $\mathbf{A}$ are generated independently and identically distributed according to a distribution that satisfies (\ref{con2:eq}), considering the model in (\ref{eq:ncs2}), we will have:
\begin{equation}
\label{eq19}
\textrm{CRB-S}\longrightarrow \alpha \sigma_n^2=\frac{K}{N} \sigma_n^2~.
\end{equation}
\end{lemma}
\begin{proof}
The proof of this lemma is given in~\cite{babadi2009asymptotic} for the special case that elements of $\mathbf{A}$ are i.i.d Gaussian random variables. Generalization of this proof for the family of distributions that satisfy (\ref{con2:eq}) is elementary and is left to the reader. 
\end{proof}

Now, we are going to investigate the relation between CRB-S and CRB-US (which is Cram\'{e}r-Rao bound of the estimators with just knowledge about the cardinality of $\tau$) under the assumption that the measurement matrix, $\mathbf{A}$, is a randomly generated but deterministic matrix that satisfies our modified concentration of measures inequality described in (\ref{con2:eq}). As was mentioned before, CRB-S and CRB-US seem to be different bounds at the first glance. But interestingly, as was shown in~\cite{babadi2009asymptotic}, in the asymptotic case they tend to each other. The proof of this statement in~\cite{babadi2009asymptotic} is based on the Lemma 3.3 in \cite{akcakaya711shannon} which is based on the Gaussianity of the measurement matrix and its randomness in the noise domain. So, one may wonder if the results in~\cite{babadi2009asymptotic} are still correct under our new generalized assumptions, which fortunately is, as we will discuss later in this section. For showing this, we investigate the method of estimation in~\cite{babadi2009asymptotic} which is based on a combinatorial search for finding the support of original sparse vector. Before proceeding any further, we will state the definition of this estimator as in~\cite{babadi2009asymptotic}:
\begin{definition}[Joint Typicality Estimator]
\label{def1}
The Joint Typicality Estimator finds a set of indices, $\zeta\subset\{1,2,\dots M\}$ with cardinality of $K$ which is jointly typical with $\mathbf{y}$ with order of $\epsilon$. After that, it will produce the estimate $\mathbf{\hat{s}}_{\zeta}$ as:
\begin{equation}
(\mathbf{A}_{\zeta}^T\mathbf{A}_{\zeta})^{-1}\mathbf{A}_{\zeta}^T\mathbf{y}~.
\end{equation} 
If the estimator does not find a unique solution for $\zeta$, it will return an all-zero vector as its output.
\end{definition}

In the main theorem of~\cite{babadi2009asymptotic}, it is shown that under certain constraints, the MSE of the jointly typical estimator is upper bounded by $\alpha\sigma_n^2$. But the proof of this property is strongly based on Lemma 3.3 in~\cite{akcakaya711shannon}, which cannot be used under our new assumptions, as was mentioned before. Instead, we use our variant of this lemma (Theorem~\ref{maint} and especially its asymptotic form in (\ref{eq18})). According to the fact that this variant and the original form in~\cite{akcakaya711shannon} are not much different from each other, we can show that the main theorem in~\cite{babadi2009asymptotic} remains valid without any necessary changes. More accurately, we have the following theorem:

\begin{theorem}[Revised version of Main Theorem in ~\cite{babadi2009asymptotic}]
\label{maint2}
Consider the model described in (\ref{eq:ncs2}) and suppose that $\mathbf{A}$ is a randomly generated, but a deterministic matrix in the noise domain that satisfies (\ref{con2:eq}). Let $\mathbf{\hat{s}}_{\zeta}$ be the output of the Jointly Typical Estimator defined in Definition~\ref{def1}. Additionally, let $\mu(\mathbf{s})\triangleq\textrm{min}_{i\in\tau}\lvert s_i\rvert$. If
\begin{itemize}
\item $\frac{K\mu^4(\mathbf{s})}{\log(K)}\rightarrow \infty$ as $N\rightarrow\infty$~,
\item $\lVert\mathbf{s}\rVert_2^2$ grows polynomially in $N$~,
\item $\alpha<\frac{1}{9+4\log(\beta-1)}$~,
\end{itemize}
then we have:
\begin{equation}
\mathbb{E}\{\lVert \mathbf{s}_{\tau}-\mathbf{\hat{s}}_{\zeta} \rVert^2\}\leq \alpha \sigma_n^2~,
\end{equation}
as $N\rightarrow\infty$ for a fixed $\alpha$ and $\beta$.
\end{theorem}
\begin{proof}
Our proof, is exactly the same as the proof in~\cite{babadi2009asymptotic} with some minor changes. First, similar to the mentioned proof, we try to upper bound the MSE of the estimation. Indeed, by repeating the first steps described by equations (17)-(22) of~\cite{babadi2009asymptotic}, applying the new form of the Lemma 3.3 in~\cite{akcakaya711shannon} which contains the bounds in (\ref{th1}) and (\ref{th2}) and also by using the asymptotic form of Theorem~\ref{maint} described in (\ref{eq18}), we can upper bound the MSE of joint typical estimator, i.e. $\mathbb{E}\{\lVert \mathbf{s}_{\tau}-\mathbf{\hat{s}}_{\zeta} \rVert^2\}$, by:
\begin{equation}
\begin{split}
\label{eq:sum}
&\alpha{\sigma'}_n^2+(K{\sigma'}_n^2+\lVert \mathbf{s}\rVert^2)\sum_{k'=1}^{K}{\binom{K}{k'}\binom{M-K}{k'}}\times\\
&\exp\Big\{-\frac{N-K}{4}\big(\frac{k'\mu^2(\mathbf{s})-\epsilon'}{2k'\mu^2(\mathbf{s})+{\sigma}_n^2}\big)^2\Big\}~.
\end{split}
\end{equation}
Similar to~\cite{babadi2009asymptotic}, we use the inequality
\begin{equation}
\binom{K}{k'}\leq\exp\Big(k'\log(\frac{Ke}{k'})\Big)~,
\end{equation}
to upper bound the $k'$-th term in the summation of (\ref{eq:sum}) by:
\begin{align}
\exp\Big( K \frac{k'}{K}\log\big(\frac{e}{\frac{k'}{K}}\big)+&K \frac{k'}{K}\log\big(\frac{(\beta-1)e}{\frac{k'}{K}}\big) -\nonumber \\
& C_0K\big(\frac{K\frac{k'}{K}\mu^2(\mathbf{s})-\epsilon'}{2K\frac{k'}{K}\mu^2(\mathbf{s})+{\sigma}_n^2}\big)^2\Big)~,
\end{align} 
in which $C_0\triangleq \frac{N-K}{4K}$. Again, similar to~\cite{babadi2009asymptotic} we define:
\begin{align}
f(z)\triangleq K z\log\big(\frac{e}{z}\big)&+K z\log\big(\frac{(\beta-1)e}{z}\big) -\nonumber\\
& C_0K\big(\frac{Kz\mu^2(\mathbf{s})-\epsilon'}{2Kz\mu^2(\mathbf{s})+{\sigma}_n^2}\big)^2~.
\end{align}
Now, by Lemmas~3.4, 3.5 and 3.6 of \cite{akcakaya711shannon} we can easily conclude that $f(z)$ attains its maximum at either $z=1$ or $z=\frac{1}{K}$ if $\frac{K\mu^4(\mathbf{s})}{\log(K)}\rightarrow \infty$ as $N\rightarrow \infty$. So, we can upper bound (\ref{eq:sum}) as:
\begin{align}
\label{eq:sum2}
&\alpha{\sigma}_n^2+(K{\sigma}_n^2+\lVert \mathbf{s}\rVert^2)\sum_{k'=1}^{K}\exp\Big\{{\max\big\{f(1),f(\frac{1}{K}) \big\}}\Big\}=\nonumber\\
&\alpha{\sigma}_n^2+\exp\Big\{\log(K^2{\sigma}_n^2+K\lVert s\rVert^2)+ \max\big\{f(1),f(\frac{1}{K})\big\} \Big\}~.
\end{align}
Additionally, we have:
\begin{equation}
f(1)= K (2+\log(\beta-1)) - C_0K\big(\frac{K\mu^2(\mathbf{s})-\epsilon'}{2K\mu^2(\mathbf{s})+{\sigma}_n^2}\big)^2~,
\end{equation}
and
\begin{equation}
f(\frac{1}{K})=2\log(K)+2+\log(\beta-1) - C_0K\big(\frac{\mu^2(\mathbf{s})-\epsilon'}{2\mu^2(\mathbf{s})+{\sigma}_n^2}\big)^2~.
\end{equation}
It is obvious that $f(\frac{1}{K})$ grows linearly to $-\infty$ as $N\to\infty$. Additionally, if $C_0>2+\log(\beta-1)$ or equivalently $\alpha<\frac{1}{9+4\log(\beta-1)}$ then $f(1)$ will also grow linearly to $-\infty$ as $N\to\infty$. Hence, the exponent of the second term in (\ref{eq:sum2}) tends to $-\infty$ as long as $\lVert \mathbf{s} \rVert^2$ grows polynomially with respect to $N$. So we have the following inequality when $N\to\infty$
\begin{equation}
\mathbb{E}\{\lVert \mathbf{s}_{\tau}-\mathbf{\hat{s}}_{\zeta} \rVert^2\}\leq \alpha \sigma_n^2~,
\end{equation}
which completes the proof.
\end{proof}

Now, by comparing the result of Theorem~\ref{maint2} with (\ref{eq19}) and (\ref{eq:neq}), we come to the conclusion that under the assumption we made about $\mathbf{A}$ (its distribution satisfies (\ref{con2:eq})), the CRB-S and CRB-US are asymptotically equal. Additionally, they can be asymptotically achieved using the Jointly Typical Estimator. 
\section{Conclusion}
In this correspondence paper, we examined the problem of the achievability of the Cram\'{e}r-Rao bound in noisy compressed sensing under some new assumptions on the measurement matrix. Indeed, we relax our analysis from the Gaussianity constraint on the measurement matrix and its randomness in the domain of noise. Instead, we assumed that this matrix is randomly generated according to a distribution that satisfies some sort of concentration of measures inequality (described in (\ref{con2:eq})), but is deterministic in the noise domain. Mainly, we focused on the proof of Lemma 3.3 in \cite{akcakaya711shannon} which was the main building block of the interesting results obtained in~\cite{babadi2009asymptotic}. After reproving a new form of the above mentioned lemma using our new assumptions, we showed that the main theorem of~\cite{babadi2009asymptotic} is still valid under these assumptions. So, the Cram\'{e}r-Rao bound of the GAE and the Cram\'{e}r-Rao bound for estimation with no prior knowledge about the original vector except for its degree of sparsity, are indeed asymptotically equal and the Jointly Typical Estimator first introduced in~\cite{babadi2009asymptotic} can achieve this bound. Unfortunately, this method of estimation is impractical and to the best knowledge of the authors, the problem of finding a practical estimator that can achieve the Cram\'{e}r-Rao bound is still open.
\begin{appendix}
\label{app}
\begin{lemma}
\label{lem1}
The function $g(\nu)$ defined in (\ref{eqmgf2}) will reach its minimum at $\nu^*$ given in (\ref{eq11}). Moreover, $\nu^*< 0$ and $1-2\sigma_n^2\nu^*>0$. 
\end{lemma}
\begin{proof}
By taking the derivative of $g(\nu)$ with respect to $\nu$ and setting it to zero, we will have the following equation for finding the roots of $\frac{\partial g}{\partial \nu}$:
\begin{equation*}
(-\bar{\epsilon}+{\sigma'}_n^{2}+\gamma^2)X^2-{\sigma'}_n^{2}X-\gamma^2=0~,
\end{equation*}
in which $X\triangleq(1-2\nu^*\sigma_n^2)$. By solving this equation with respect to $X$, we will have two solutions for $X$:
\begin{equation*}
X_1=\frac{{\sigma'}_n^{2}+\sqrt{({\sigma'}_n^{2}+2\gamma^2)^2-4\gamma^2\bar{\epsilon}}}{2(-\bar{\epsilon}+{\sigma'}_n^{2}+\gamma^2)}~,
\end{equation*} 
and
\begin{equation*}
X_2=\frac{{\sigma'}_n^{2}-\sqrt{({\sigma'}_n^{2}+2\gamma^2)^2-4\gamma^2\bar{\epsilon}}}{2(-\bar{\epsilon}+{\sigma'}_n^{2}+\gamma^2)}~.
\end{equation*} 
First, it is important to note that both of these solutions are real, as by substituting $\gamma^2-\bar{\epsilon}$ with $\epsilon$ we will have that $({\sigma'}_n^{2}+2\gamma^2)^2-4\gamma^2\bar{\epsilon}={\sigma'}_n^4+4\gamma^2{\sigma'}_n^2+4\gamma^2\epsilon>0$. Furthermore, it also shows that $X_1>0$ and $X_2<0$. As we are looking for a $\nu^*$ that satisfies the constraint $1-2\nu^*\sigma_n^2>0$, the latter solution $X_2$ is not acceptable, and so we have $X=(1-2\nu^*\sigma_n^2)=X_1$. By taking the second derivative of $g(\nu)$ with respect to $\nu$, it is easy to show that $\frac{\partial^2}{\partial \nu^2}g(\nu^*)\geq 0$ and so $g(\nu)$ will reach its minimum value at $\nu^*$. It is important to note that the following expressions are also valid and can be extracted from the expression for $X$:
\begin{align}
&1-2\nu^*\sigma_n^2=\frac{{\sigma'}_n^{2}+\sqrt{({\sigma'}_n^{2}+2\gamma^2)^2-4\gamma^2\bar{\epsilon}}}{-2(\bar{\epsilon}+{\sigma'}_n^{2}+\gamma^2)}\\
&\frac{1}{1-2\nu^*\sigma_n^2}=\frac{\sqrt{({\sigma'}_n^{2}+2\gamma^2)^2-4\gamma^2\bar{\epsilon}}-{\sigma'}_n^{2}}{2\gamma^2}\\
&\nu^*=\frac{2\gamma^2-2\bar{\epsilon}+{\sigma'}_n^{2}-\sqrt{({\sigma'}_n^{2}+2\gamma^2)^2-4\gamma^2\bar{\epsilon}}}{4\sigma_n^{2}(\gamma^2-\bar{\epsilon}+{\sigma'}_n^{2})}\\
&\label{eqfourth}\frac{\gamma^2\nu^*}{1-2\nu^*\sigma_n^2}=\frac{\sqrt{({\sigma'}_n^{2}+2\gamma^2)^2-4\gamma^2\bar{\epsilon}}-{\sigma'}_n^{2}-2\gamma^2}{4\sigma_n^2}~.
\end{align}    
By looking at (\ref{eqfourth}), it is obvious that the nominator of the right hand side of this equation is a negative term (because $\sqrt{({\sigma'}_n^{2}+2\gamma^2)^2-4\gamma^2\bar{\epsilon}}<{\sigma'}_n^{2}+2\gamma^2$), while the denominator, i.e. $4\sigma_n^2$, is a positive term. So, by using the fact that $1-2\sigma_n^2\nu^*>0$ (as we have proven before), one can concludes that $\nu^*<0$, which completes the proof.
\end{proof}
\end{appendix}
\section*{Acknowledgment}
The authors would like to thank Maryam Sharifzadeh, MS student of Mathematical and Computer Science Department at Sharif University of Technology, for her helpful comments. Additionally, the authors thank Prof. Vahid Tarokh for his remarkable research around this subject which was the main motivation for this work. The authors are also grateful to the anonymous reviewers and to the associate
editor for their constructive comments.

\bibliographystyle{IEEEtran}
\bibliography{refs}

\begin{thebibliography}{10}
\providecommand{\url}[1]{#1}
\csname url@samestyle\endcsname
\providecommand{\newblock}{\relax}
\providecommand{\bibinfo}[2]{#2}
\providecommand{\BIBentrySTDinterwordspacing}{\spaceskip=0pt\relax}
\providecommand{\BIBentryALTinterwordstretchfactor}{4}
\providecommand{\BIBentryALTinterwordspacing}{\spaceskip=\fontdimen2\font plus
\BIBentryALTinterwordstretchfactor\fontdimen3\font minus
  \fontdimen4\font\relax}
\providecommand{\BIBforeignlanguage}[2]{{%
\expandafter\ifx\csname l@#1\endcsname\relax
\typeout{** WARNING: IEEEtran.bst: No hyphenation pattern has been}%
\typeout{** loaded for the language `#1'. Using the pattern for}%
\typeout{** the default language instead.}%
\else
\language=\csname l@#1\endcsname
\fi
#2}}
\providecommand{\BIBdecl}{\relax}
\BIBdecl

\bibitem{babadi2009asymptotic}
B.~Babadi, N.~Kalouptsidis, and V.~Tarokh, ``{Asymptotic achievability of the
  Cram{\'e}r--Rao bound for noisy compressive sampling},'' \emph{IEEE
  Transactions on Signal Processing}, vol.~57, no.~3, pp. 1233--1236, 2009.

\bibitem{akcakaya711shannon}
M.~Ak{\c{c}}akaya and V.~Tarokh, ``{Shannon theoretic limits on noisy
  compressive sampling},'' \emph{IEEE Transactions on Information Theory},
  vol.~56, no.~1, pp. 492--504, 2010.

\bibitem{donoho2006compressed}
D.~Donoho, ``{Compressed sensing},'' \emph{IEEE Transactions on Information
  Theory}, vol.~52, no.~4, pp. 1289--1306, 2006.

\bibitem{candes2006compressive}
E.~Cand{\`e}s, ``{Compressive sampling},'' in \emph{Proceedings of the
  International Congress of Mathematicians}, vol.~3.\hskip 1em plus 0.5em minus
  0.4em\relax Citeseer, 2006, p. 14331452.

\bibitem{tsaig2006extensions}
Y.~Tsaig and D.~Donoho, ``{Extensions of compressed sensing},'' \emph{Signal
  processing}, vol.~86, no.~3, pp. 549--571, 2006.

\bibitem{candes2006robust}
E.~Cand{\`e}s, J.~Romberg, and T.~Tao, ``{Robust uncertainty principles: Exact
  signal reconstruction from highly incomplete frequency information},''
  \emph{IEEE Transactions on information theory}, vol.~52, no.~2, p. 489, 2006.

\bibitem{baraniuk2008compressive}
R.~Baraniuk, E.~Cand{\`e}s, R.~Nowak, and M.~Vetterli, ``{Compressive
  sampling},'' \emph{IEEE Signal Processing Magazine}, vol.~25, no.~2, pp.
  12--13, 2008.

\bibitem{niazadeh2010alternating}
R.~Niazadeh, M.~Babaie-Zadeh, and C.~Jutten, ``{An alternating minimization
  method for sparse channel estimation},'' in \emph{Ninth International
  Conference on Latent Variable Analysis and Signal Seperation (LVA-ICA,
  formerly known as ICA)}.\hskip 1em plus 0.5em minus 0.4em\relax
  Springer-Verlag, 2010, pp. 319--327.

\bibitem{candes2006stable}
E.~Cand{\`e}s, J.~Romberg, and T.~Tao, ``{Stable signal recovery from
  incomplete and inaccurate measurements},'' \emph{Communications on Pure and
  Applied Mathematics}, vol.~59, no.~8, pp. 1207--1223, 2006.

\bibitem{tropp2007signal}
J.~Tropp and A.~Gilbert, ``{Signal recovery from random measurements via
  orthogonal matching pursuit},'' \emph{IEEE Transactions on Information
  Theory}, vol.~53, no.~12, p. 4655, 2007.

\bibitem{donoho2006most}
D.~Donoho, ``{For most large underdetermined systems of linear equations the
  minimal l1-norm solution is also the sparsest solution},''
  \emph{Communications on Pure and Applied Mathematics}, vol.~59, no.~6, pp.
  797--829, 2006.

\bibitem{chen2001atomic}
S.~Chen, D.~Donoho, and M.~Saunders, ``{Atomic decomposition by basis
  pursuit},'' \emph{SIAM review}, vol.~43, no.~1, pp. 129--159, 2001.

\bibitem{blumensath2008gradient}
T.~Blumensath and M.~Davies, ``{Gradient pursuits},'' \emph{IEEE Transactions
  on Signal Processing}, vol.~56, no.~6, pp. 2370--2382, 2008.

\bibitem{gorodnitsky1997sparse}
I.~Gorodnitsky and B.~Rao, ``{Sparse signal reconstruction from limited data
  using FOCUSS: are-weighted minimum norm algorithm},'' \emph{IEEE Transactions
  on Signal Processing}, vol.~45, no.~3, pp. 600--616, 1997.

\bibitem{needell2009cosamp}
D.~Needell and J.~Tropp, ``{CoSaMP: Iterative signal recovery from incomplete
  and inaccurate samples},'' \emph{Applied and Computational Harmonic
  Analysis}, vol.~26, no.~3, pp. 301--321, 2009.

\bibitem{mohimani2009fast}
H.~Mohimani, M.~Babaie-Zadeh, and C.~Jutten, ``{A Fast Approach for
  Overcomplete Sparse Decomposition Based on Smoothed l0 Norm},'' \emph{IEEE
  Transactions on Signal Processing}, vol.~57, no.~1, pp. 289--301, 2009.

\bibitem{kay1993fundamentals}
S.~Kay, ``{Fundamentals of statistical signal processing: estimation theory},''
  \emph{Prentice-Hall Signal Processing Series}, 1993.

\bibitem{carbonelli2007sparse}
C.~Carbonelli, S.~Vedantam, and U.~Mitra, ``{Sparse channel estimation with
  zero tap detection},'' \emph{IEEE Transactions on Wireless Communications},
  vol.~6, no.~5, pp. 1743--1763, 2007.

\bibitem{candes2007dantzig}
E.~Cand{\`e}s and T.~Tao, ``{The Dantzig selector: statistical estimation when
  p is much larger than n},'' \emph{Annals of Statistics}, vol.~35, no.~6, pp.
  2313--2351, 2007.

\bibitem{ben2010cramér}
Z.~Ben-Haim and Y.~Eldar, ``{The Cram{\'e}r-Rao bound for estimating a sparse
  parameter vector},'' \emph{Signal Processing, IEEE Transactions on}, vol.~58,
  no.~6, pp. 3384--3389, 2010.

\bibitem{haupt2006signal}
J.~Haupt and R.~Nowak, ``{Signal reconstruction from noisy random
  projections},'' \emph{IEEE Transactions on Information Theory}, vol.~52,
  no.~9, pp. 4036--4048, 2006.

\bibitem{cover2006elements}
T.~Cover and J.~Thomas, \emph{{Elements of information theory}}.\hskip 1em plus
  0.5em minus 0.4em\relax Wiley, 2006.

\bibitem{shannon2001mathematical}
C.~Shannon, ``{A mathematical theory of communication},'' \emph{ACM SIGMOBILE
  Mobile Computing and Communications Review}, vol.~5, no.~1, p.~55, 2001.

\bibitem{baraniuk2008simple}
R.~Baraniuk, M.~Davenport, R.~DeVore, and M.~Wakin, ``{A simple proof of the
  restricted isometry property for random matrices},'' \emph{Constructive
  Approximation}, vol.~28, no.~3, pp. 253--263, 2008.

\bibitem{achlioptas2003database}
D.~Achlioptas, ``{Database-friendly random projections: Johnson-Lindenstrauss
  with binary coins},'' \emph{Journal of Computer and System Sciences},
  vol.~66, no.~4, pp. 671--687, 2003.

\bibitem{hagerup1990guided}
T.~Hagerup and C.~Rub, ``{A guided tour of Chernoff bounds},'' \emph{Proc.
  Lett}, vol.~33, pp. 305--308, 1989.

\bibitem{papoulis1965probability}
A.~Papoulis, S.~Pillai, P.~A, and P.~SU, \emph{{Probability, random variables,
  and stochastic processes}}.\hskip 1em plus 0.5em minus 0.4em\relax
  McGraw-Hill New York, 1965.

\bibitem{haupt2010toeplitz}
J.~Haupt, W.~Bajwa, G.~Raz, and R.~Nowak, ``{Toeplitz compressed sensing
  matrices with applications to sparse channel estimation},'' \emph{Information
  Theory, IEEE Transactions on}, vol.~56, no.~11, pp. 5862--5875, 2010.

\end{thebibliography}
\end{document}